\documentstyle[aps,prl,multicol,psfig]{revtex}
\begin{document}
\draft

\title{Influence of Capillary Condensation on the Near-Critical Solvation Force}
\author{A. Drzewi\'nski$^1$,  A. Macio\l ek$^{2}$, and  R. Evans$^3$}
\address{$^1$Institute of Low Temperature and Structure Research,
Polish Academy of Sciences, P.O.Box 1410, 50-950 Wroc\l aw 2, Poland \\
$^{2}$Institute of Physical Chemistry, Polish Academy of Sciences,
Department III, Kasprzaka 44/52, PL-01-224 Warsaw, Poland \\
$^3$H.H.Wills Physics Laboratory, University of  Bristol, 
Bristol BS8 1TL, UK }
\date{\today}
\maketitle
            
\begin{abstract}
We argue that in a fluid, or magnet, confined by  adsorbing  walls which favour 
liquid, or $(+)$ phase, the solvation (Casimir) force in 
the vicinity of the critical point is strongly influenced
 by capillary condensation which occurs below the bulk critical temperature 
$T_c$. At $T$ slightly below and above $T_c$, a small  bulk field
$h<0$, which favours gas, or $(-)$ phase, leads to residual condensation
and a solvation force which is much more attractive (at the same large wall 
separation) than  that found exactly at the  critical point. Our predictions are 
supported by results obtained from density-matrix renormalization-group calculations 
in a two-dimensional Ising strip subject to identical surface fields.
\end{abstract}

\begin{multicols}{2} \narrowtext
Finite-size contributions to the free energy of a fluid confined between two 
parallel walls, separated by a distance $L$,
give rise to a force per unit area between the walls or an excess pressure
which is termed the solvation force $f_{solv}(L)$~\cite{evans:90:0}.
This force is, essentially, that which is  measured in the surface force 
apparatus~\cite{israelachvili:91:0}.
Theory predicts that at the critical point of a fluid the solvation force becomes 
long ranged as a result of critical
fluctuations~\cite{FdeG:78:0}, a phenomenon  which is a direct analog of 
the well-known Casimir effect in electromagnetism~\cite{casimir:48:0}.
The existence of the long-ranged critical Casimir force should be 
common to all systems characterized by fluctuating quantities with external
constraints~\cite{krech:94:0}. As yet there has been no direct, unambiguous 
experimental verification of the critical
Casimir effect in fluids~\cite{krech:94:0}, although recent experiments do 
provide indirect evidence for its existence~\cite{law:99:0}. 
One of the difficulties is that the predicted leading  power law decay 
of the Casimir force at bulk criticality, $f_{solv}(L)\sim 
k_BT_cA_{12}(d-1)L^{-d}$, as $L\to \infty$, is for (bulk) spatial dimension $d=3$, of
 the same form as the solvation force arising from dispersion forces. Moreover  
the amplitude in many systems may be much smaller than 
the corresponding Hamaker constant~\cite{krech:94:0,krech:91:0}. 
The Casimir amplitude $A_{12}$ is a universal number, however, its value 
depends on the type of boundary conditions imposed on the two walls~\cite{krech:94:0}.
The most relevant  for  experiments on pure fluids
or for binary mixtures are  symmetry breaking boundary conditions, i.e.,
when the confining  walls exert  surface fields on molecules in the fluid.
Here we exploit the mapping between  fluids and the Ising model and
consider Ising spin systems subject to identical surface fields $h_1=h_2>0$.
For these systems $f_{solv}(L)$ is expected to be attractive for all 
thermodynamic states. 
In $d=2$, the Casimir amplitude is known exactly $A_{11}\equiv 
A_{++}=-\pi/48$~\cite{blote:86:0},
whereas in $d=3$ the most recent theoretical result 
is -0.428~\cite{borjan:99:0} and the Monte Carlo estimate is 
-0.35~\cite{krech:97:0}.
In this Letter we show that other, near-critical, thermodynamic states exhibit a  
significantly stronger solvation force (for fixed, large $L$) than that found
exactly at the bulk critical point, and we suggest that this has repercussions 
for experimental studies.

For Ising-like systems with  $h_1=h_2>0$ in vanishing
 bulk field $h=0$, $ f_{solv}$  as a 
function of temperature
attains a pronounced minimum 
{\it  above} the bulk critical temperature $T_c$.
The scaling function $W_{++}(y)$, defined by $f_{solv}/k_BT_c\equiv 
L^{-d}W_{++}(y)$, 
was determined by exact transfer matrix methods in $d=2$~\cite{evans:93:0}.
The minimum occurs when $y\equiv \tau(L/\xi^+_0)^{1/\nu}=2.23$, or $L\sim 2.23 \xi$, and the amplitude
 at this extremum is about 6.6 times the Casimir value $W_{++}(0)=A_{++}(d-1)$.
Here $\tau\equiv (T-T_c)/T_c$ and $\xi=\xi^+_0\tau^{-\nu}$ (with $\nu=1$ in 
$d=2$) is the  bulk correlation length.
In mean field the minimum occurs for $L\sim 3.7 \xi$ and the amplitude
 is about 1.4 times $W_{++}(0)$~\cite{krech:97:0,hanke:98:0}.
Borjan and Upton~\cite{zoran:99:0} have calculated  $W_{++}(y)$ in $d=3$ using local functional
methods and they find the minimum at  $L\sim 3.1\xi $, and the 
amplitude at the
extremum is about 2.1 times $W_{++}(0)$.
The bulk field dependence of the solvation force at $T=T_c$
is  similar to the 
temperature dependence at $h=0$, i.e., $ f_{solv}$ attains a pronounced
 minimum at some  $h_{min}<0$~\cite{drzew:00:0,frank:00:0}.
In $d=2$ the scaling function of $f_{solv}$ with $h$  was obtained 
using the density-matrix renormalization-group (DMRG)
method. The striking feature is that at the minimum this function  is about 100 times  bigger than the 
Casimir amplitude~\cite{drzew:00:0}!
In mean field it is about 10 times  the Casimir value~\cite{frank:00:0}.
Thus, the solvation force is much stronger for states which lie slightly 
 {\it off  bulk coexistence}, with $h<0$.
 In this Letter we investigate the origin of the very attractive solvation force  
and  argue that  in a fluid confined by identical, 
strongly adsorbing  walls 
the form of the near-critical, 
long-ranged solvation force is strongly influenced
 by capillary condensation, i.e., the shift of the bulk first-order
transition which occurs {\it below} $T_c$.
Recall that for surface fields $h_1=h_2>0$ and bulk dimension $d\ge 3$,
two phase coexistence occurs in the confined system, with finite (large) $L$, 
along
a line $h_{co}(T)$ ending in a (capillary) critical point 
$(h_{CL}, T_{CL})$. $T_{CL}(h_1)$ lies below $T_C$ and 
$h_{CL}(h_1)<0$~\cite{fisher:81:0,evans:86:0}.
The coexistence (capillary condensation) line  has a positive slope and is 
located at $h<0$.
$f_{solv}$ exhibits a  discontinuous jump on crossing the coexistence line and 
singular behaviour at 
$(h_{CL},T_{CL})$~\cite{evans:90:0,evans:86:0};
 for $d\ge 3$ the capillary criticality corresponds to the $d-1 $ universality 
class.
The form of  $f_{solv}$ for $T>T_{CL}$ and $h<0$ does not seem to have been 
investigated  in any detail.
We perform such an investigation for  Ising films using the DMRG method.
Although for bulk $d=2$ systems  there can be no  true phase transition for 
finite $L$,
 there is still a line of sharp (very weakly rounded) first-order transitions in 
the Ising film 
ending in a pseudocritical point~\cite{note:0,albano:89:0,carlon:98:0}.
We  determine this pseudocoexistence line and investigate $f_{solv}$ for 
temperatures 
above the capillary condensation line
and states approaching the bulk critical regime. 

\begin{figure}[b]
\centerline{
\psfig{file=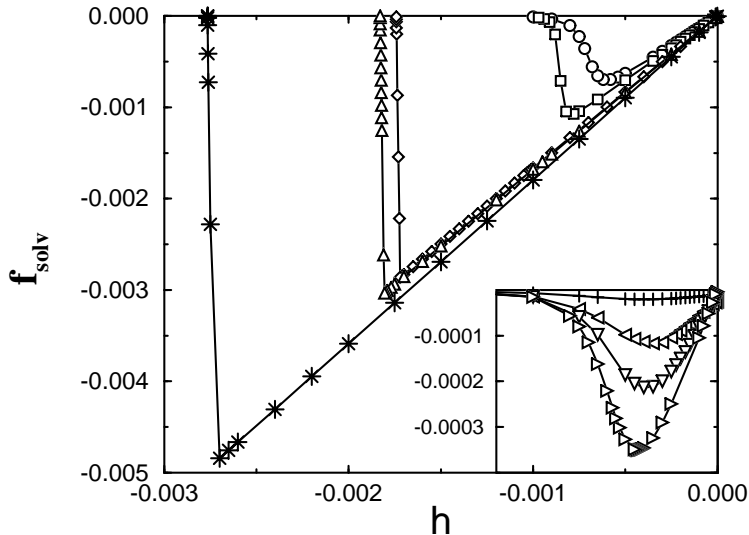,height=8.cm}}
\vskip 0.2truecm
\caption{Solvation force 
as a function of the bulk magnetic field $h$ (both quantities in units of $J$) for an
Ising strip of width $L=200$, surface fields $h_1=h_2=10$ and several reduced 
temperatures $\tau \equiv (T-T_c)/T_c \approx $: (+) 0.0135, ($\triangleleft$) 0.0063, 
($\bigtriangledown$) 0.0031, ($\triangleright$) 0.0, ($\circ$) -0.0063, ($\Box$) -0.0126,
($\diamond$) -0.0505, ($\bigtriangleup$) -0.0546, and ($\ast $) -0.1011.}
\label{fig:fsol}
\end{figure}

\begin{figure}[b]
\centerline{
\psfig{file=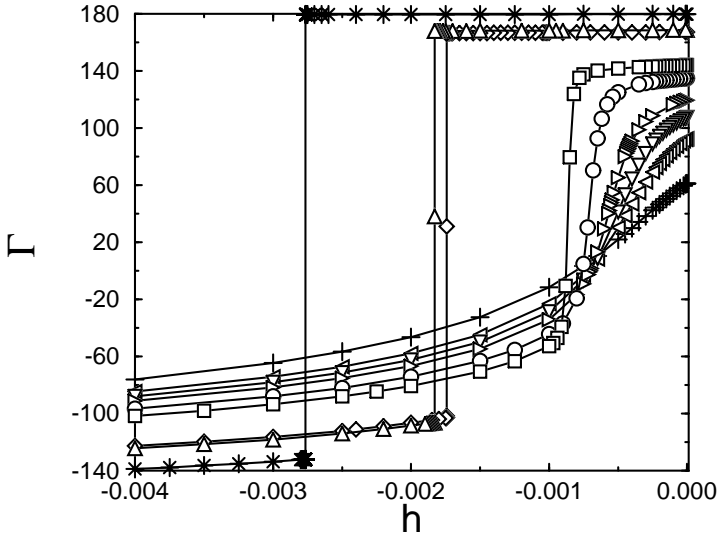,height=8.cm}}
\vskip 0.2truecm
\caption{Total adsorption $\Gamma$ vs $h$ 
(in units of $J$) for the same system and the same reduced temperatures $\tau$ 
as in Fig.~\ref{fig:fsol}.
}
\label{fig:ads}
\end{figure}

Consider a $d=2$ Ising strip with $L$ parallel rows  subject to surface fields 
$h_1=h_2(\equiv h_L)>0$ 
with the Hamiltonian $H$:
\begin{equation}
\label{eq:ham}
-H/J = \displaystyle{\sum _{<i,j>}}\sigma _i\sigma _j
             +h\sum _i\sigma _i
            +h_1\sum _i^{(1)}\sigma _i+h_L\sum _i^{(L)}\sigma _i,
\end{equation}
where  $\sigma _i$ is a spin variable taking the value $\pm 1$ and $J>0$ is the 
coupling constant.
The first sum runs over all nearest-neighbour pairs
 of sites while the last
two sums run, respectively, over the first and the $L$th row. $h$ is the bulk 
magnetic field.
The total free energy per site of this system can be written as
$f(L,T,h_1,h)=f_b(T,h)+\frac{2}{L}f_w(T,h_1,h)+\frac{1}{L}f^*(L,T,h_1,h)$,
where $f_b$ is the bulk free energy, $f_w$ is the $L$-independent
wall tension at each wall, and $f^*(L)$ is the finite-size 
contribution to the free energy, which vanishes for $L\to \infty$.
All energies are measured in units of $J$, temperature in units of $J/k_B$ and 
the width $L$ in units
of the lattice constant $a$. The solvation force for our system is defined as
$f_{solv}=-(\partial f^{*}/\partial L)_{T,h_1,h}$.
We calculate $f_{solv}$  using the
recently developed, very accurate  DMGR method, which 
 provides an efficient algorithm for the construction of the effective 
transfer
matrix for large systems~\cite{nishino:95:0}. 
A great advantage of this method is that it  works equally well for $h\ne 0$,
for which no exact solutions are available, as for $h=0$. The total free energy 
$f$ is obtained from the leading eigenvalue $\Lambda _0$
of the effective transfer matrix which in the DMRG method  
is calculated numerically~\cite{drzew:00:0}.
By calculating the excess free energy per unit area $ f_{ex}(L)\equiv 
\left(f-f_b\right)L$ at $L_0+2$ and 
 $ L_0$ we obtain $f_{solv}$  by a finite difference. 
 Figure \ref{fig:fsol} summarizes our results for $f_{solv}$ in a film of  width 
$L=200$
and $h_1=10$, conditions which pertain to the infinite surface field scaling 
limit~\cite{drzew:00:0}.
$f_{solv}$ is calculated as a function of $h$ for several 
temperatures above and below $T_c$.
For the three lowest  temperatures shown
we find a very weakly rounded jump of
the solvation force from zero to a negative value. As mentioned earlier, 
a discontinuous jump is  characteristic  of the solvation force at a first-order 
capillary condensation 
phase transition~\cite{evans:90:0,evans:86:0}.
At fixed large $L$ and fixed temperature $T<T_{CL}$, $f_{solv}$ should  change 
abruptly from 
values appropriate to a
spin down $(-)$: $f_{solv}^- \approx 0$ to values appropriate to a spin up 
$(+)$: $f_{solv}^+\approx 2hm^*(T)$,
where $m^*(T)>0$ is the bulk spontaneous magnetization~\cite{evans:87:0}.
The calculated gradients of $f_{solv}$ at the three lowest temperatures in Fig.~\ref{fig:fsol} agree with the 
known values of $2m^*(T)$ to a relative accuracy of $10^{-4}$.
 Coexistence is given by the Kelvin equation~\cite{evans:87:0}
 which for the strong surface fields $h_1$ considered here takes the form
$-h_{co}(T)\approx \sigma (T)/Lm^*(T)$, where $\sigma (T)$ 
is the interfacial tension between coexisting (+) and (-) phases.
 The presence of thick wetting films 
of  + spin in the (-) phase 
gives rise to non-trivial corrections which shift the condensation line 
to larger values of $\mid h \mid $~\cite{parry:92:0}; nevertheless the Kelvin 
equation does predict the correct qualitative
behavior of the condensation line at low temperatures. The Kelvin result implies
$\Delta f_{solv}\equiv f_{solv}^{+}-f_{solv}^{-}\approx 2h_{co}(T)m^*(T)\approx 
-2\sigma (T)/L$, i.e., the magnitude of the jump should decrease in the same 
fashion as the interfacial tension, as $T$ 
increases at fixed $L$.
Our numerical results confirm the predicted trend~\cite{note:1}.
A second signature of capillary condensation is a discontinuous jump (see Fig.~\ref{fig:ads}) of the total adsorption,
$\Gamma =\sum _{l=1}^L<\sigma _l>$, from negative values characteristic of a (-) phase, 
with wetting films of + spin, 
to positive values  characteristic of a (+) phase condensing in the slit.
The  simple formula  
$\Delta \Gamma\equiv \Gamma ^+-\Gamma ^-\approx 2Lm^*(T)$~\cite{evans:87:0}
provides a good qualitative description of the jump
$\Delta \Gamma$ obtained by DMRG for the three lowest temperatures.

\begin{figure}[b]
\centerline{
\psfig{file=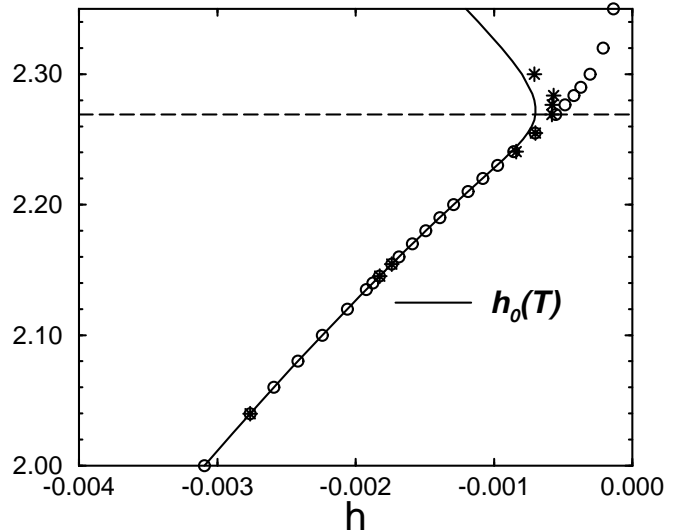,height=8.cm}}
\vskip 0.2truecm
\caption{Maxima $h_0(T)$ of the free energy 
(or zeros of the adsorption $\Gamma $) (solid line), inflection points of 
$f_{solv}$ ($\ast $) and minima of $\xi ^{-1}_{\parallel}$ ($\circ $) as 
functions of the bulk magnetic field $h$ (in units of $J$) calculated at 
fixed temperature $T$ (in units of $J/k_B$) for the same system as in 
Figs. ~\ref{fig:fsol}~and~\ref{fig:ads}. The bulk critical temperature 
$T_c\approx 2.269185$ ($\tau=0$) is denoted by the horizontal line. 
Pseudocoexistence occurs along $h_0(T)$ for $T \mathrel{<\atop\sim} 2.16$.}
\label{fig:coex}
\end{figure}

The jumps in $\Gamma $ are slightly  less rounded than in $f_{solv}$  but they 
occur at the same value of $h$. This is illustrated further in Fig.~\ref{fig:coex} where 
for each $T$ we plot the inflection point $(\ast)$  of  $f_{solv}$. Note that 
these points lie on the 
line $h_0(T)$ defined by $\Gamma =0$ or, equivalently, by the maxima of the 
total free energy $f$.
For temperatures higher than $\tau=-0.0505$ 
we observe a change in the behaviour of $f_{solv}$ and $\Gamma$; see 
Figs.~\ref{fig:fsol} and~\ref{fig:ads}. 
As the  temperature increases the jump  of $f_{solv}$ gradually transforms 
into a minimum, whose depth decreases and moves monotonically towards $h=0$.
$\tau =0.0063$ corresponds to the temperature at which $ f_{solv}$
exhibits a minimum as a function of $\tau$ for $h=0$ and $L=200$ ~\cite{evans:93:0}.
At this temperature the minimum  of $f_{solv}$ as a function of $h$ is still 
present and its depth is about 4.8  times bigger than $\mid f_{solv}(h=0)\mid$.
For temperatures up to the critical temperature the magnetization profiles for 
small $\mid h\mid$ are similar
to that of a ``condensed'' (+) (liquid) phase. Moreover,  $f_{solv}$ still varies 
linearly with $h$, as is implied
by the approximate treatment above, although the linear region does shrink as 
$\tau \to 0^-$. 
The adsorption $\Gamma $ exhibits a steep increase at $\tau=0$ (Fig.~\ref{fig:ads}) 
indicating  residual  condensation.
However, the locus of $h_0(T)$ moves to larger $\mid h \mid$ for $\tau >0$, 
as does the inflection point of $f_{solv}$; see Fig.~\ref{fig:coex}.
The longitudinal spin-spin
correlation  length $\xi _{\parallel}$ provides an alternative signature of 
capillary condensation.
 If we take the transfer matrix 
in the infinite dimension then
$\xi _{\parallel}^{-1}(L,T,h_1,h)=-\ln [\Lambda_1/\Lambda_0]$,
where $\Lambda_0$ and $\Lambda_1$ are the largest and the second largest 
eigenvalues.
We  calculated $\xi_{\parallel}^{-1}$ as a function of $h$ for a series of 
temperatures.
At  the three lowest temperatures 
considered in Figs.~\ref{fig:fsol}~and~\ref{fig:ads}, where 
 the solvation force and the adsorption behave in a way characteristic 
of capillary condensation,   $\xi_{\parallel}^{-1}$ has a sharp  minimum at some 
$h_{min}(T)<0$
with  $\xi_{\parallel}^{-1}(h_{min})\approx 0$. The values of $h_{min}(T)$ lie 
very close to those where
$f_{solv}$  jumps and to $h_0(T)$; see Fig.~\ref{fig:coex}.
Recall that in Ising strips with  periodic boundary conditions
the two largest eigenvalues of the transfer matrix are asymptotically degenerate
such that, as $L\to \infty$, $\xi_{\parallel}^{-1}\sim \exp (-L\sigma (T)/k_BT)$ 
for states on the line of pseudocoexistence
 $h=0$, $T<T_c$.
Moreover, as in the periodic system ~\cite{abraham:95:0}, our calculated $\xi 
_{\parallel}^{-1}$ is symmetric and 
increasing linearly in $\mid h\mid$
  in the close neighbourhood  of its minimum, i.e., near pseudocoexistence. 
As the temperature increases the minimum gradually lifts away from zero and
becomes more shallow. Its location remains close to  $h_0(T)$ until just below 
$T_c$. For $T>T_c$, $h_{min}(T)$
moves towards  $h=0$; see Fig.~\ref{fig:coex}~\cite{note:2}. 

In the light of these various results it is natural to identify $h_0(T)$, at 
sufficiently  low $T$, 
 with the pseudocoexistence  line $h_{co}(T)$. 
Determining the pseudocritical temperature $T_{CL}$ is more difficult. However, 
different criteria, namely
the erosion of the jumps in $f_{solv}$ and in  $\Gamma $ and the lifting away 
from zero of the minimum of 
$\xi_{\parallel}^{-1}$, all yield similar estimates. We also
calculated the average ``central bond energy'' 
$U_{L/2}\equiv <\sigma _{L/2}\sigma _{L/2+1}>$  at points on the line $h_0(T)$ and
 $\partial U_{L/2}/\partial T$ on  that line. 
The latter has a pronounced minimum at $T\approx 2.145$  where $h_0(T)\approx 
-0.00184$ for the case $L=200, h_1=10$ considered here.
Although the critical point is not sharp in this system the various criteria do 
indicate a significant change in character
for $T$ between 2.155 and 2.160, providing the estimate for $T_{CL}$. Such estimates are consistent
with the scaling relations~\cite{fisher:81:0}  for 
$(h_{CL},T_{CL})$~\cite{note:2}.

We expect that in $d=3$, where there is true coexistence for 
$T<T_{CL}$, $~f_{solv}$ and $\Gamma $ should behave in a similar fashion 
to what is observed in $d=2$, but now the jumps of $f_{solv}$ and $\Gamma $ will be
discontinuous. 
Also on the critical isotherm $T=T_{CL}$, $\Delta \Gamma \sim 
(h_{CL}-h)^{1/\delta}$, as $h\to h_{CL}^-$, 
and $\delta=15$, the $d=2$ Ising exponent. 
Explicit mean-field results for a Landau (square-gradient) theory support our 
expectations for $T<T_{CL}$ and for $T>T_{CL}$, where $f_{solv}\approx 2hm^*(T)$ for $h\to 0^-$~\cite{frank:00:0}. 
Our results imply that the solvation force in a real confined fluid at temperatures near $T_c$
and at reservoir densities slightly below the critical value (or compositions 
slightly away from the critical composition in a binary mixture) should be
much more attractive than the Casimir value for the same $L$, although
the effect may be less pronounced than in $d=2$. 
In keeping with others~\cite{krech:94:0,krech:97:0,hanke:98:0} we suggest that 
experiments which aim to measure the Casimir force, e.g. by atomic force microscopy
or surface force apparatus, and future theoretical work 
should focus on the $\tau$ and $h$  dependence 
of $f_{solv}$, i.e., the scaling function, not just the Casimir amplitude.
Finally we note that the existence of a long-ranged, strongly attractive solvation force 
between two large colloidal particles immersed in a near-critical fluid can have ramifications
for aggregation or flocculation of the particles~\cite{hanke:98:0}. Studies such as the present
 should indicate where aggregation is potentially the strongest.
\acknowledgments
We are grateful to  A. Ciach and P. J. Upton for helpful discussions 
and F. Schlesener for 
informing us of his results.
 This work was partially funded by KBN grants No. 2P03B10616 and
 No. 3T09A07316 and by the Polish Academy of Sciences - Royal Society Travel Grants.

\end{multicols}


\begin{thebibliography}{99}
\bibitem{note:1} The interfacial tension in the $d=2$ Ising model is given 
exactly by $\sigma (T)=2k_BT(K-K^*)$,
where $K=J/k_BT$ and $K^*$ satisfies $\sinh (2K)\sinh (2K^*)=1$.
\bibitem{note:2} Detailed results for $\xi _{\parallel}^{-1}$, $U_{L/2}$ and for
the location of the pseudocapillary critical point and its scaling with $L$, 
for $h_1=10$ and $h_1=0$,
will be given elsewhere.
\end{thebibliography}
\end{document}